\documentclass[twocolumn,showpacs,aps,prd]{revtex4}
\pdfoutput=1
\newcommand{\revision}[1]{{           {#1}}} 
\newcommand{\half}{\frac{1}{2}}
\newcommand{\thrd}{\frac{1}{3}}

\newcommand{\sxth}{\frac{1}{6}}
\newcommand{\dar}{\partial_r}
\newcommand{\R}{{\bar r}}
\begin{document}
\title{Weyl conformal symmetry model of the dark galactic halo} 
\author{R. K. Nesbet }
\affiliation{
IBM Almaden Research Center,
650 Harry Road,
San Jose, CA 95120-6099, USA
\begin{center}rkn@earthlink.net\end{center}
}
\date{\today}
\begin{abstract}
The postulate of universal conformal (local Weyl scaling) symmetry 
modifies both general relativity and the Higgs scalar field model.
The conformal Higgs model (CHM) generates an effective cosmological 
constant that fits observed accelerating Hubble expansion   
for redshifts $z\leq 1$ (7.33 Gyr) accurately 
with only one free parameter.  Growth of a galaxy is modeled by central
accumulation of matter from an enclosing empty spherical halo 
whose radius expands with depletion.  Details of this process account 
for the nonclassical radial centripetal acceleration observed as 
excessive orbital velocities in galactic haloes.  There is no need for 
dark matter.
\end{abstract}
\maketitle
\section{Introduction}
\par Universal conformal symmetry, requiring local Weyl scaling
covariance\cite{WEY18,WEY18a,MAN06,MAN07} of all elementary physical 
fields\cite{NES13}, offers a paradigm alternative to consensus 
$\Lambda CDM$ for cosmology,  
motivated by the absence of experimental confirmation of conjectured 
dark matter and need for explanation of currently accelerating Hubble 
expansion. The conformal Higgs model (CHM)\cite{NESM1,NESM2,NESM3,NESM5}
retains the Higgs mechanism for gauge boson mass, but acquires a
gravitational term in the scalar field Lagrangian density.  The CHM
determines centrifugal cosmic acceleration accurately for redshifts 
$z\leq 1$ (7.33 Gyr)\cite{NESM1,NESM2}. Conformal gravity (CG) replaces 
the Einstein Lagrangian density by a quadratic contraction of the 
conformal Weyl tensor \cite{MAK89,MAK94,MAN90,MAN91,MAN06,MAN12}.
Substantial empirical support for this proposed break with convention 
is provided by applications of CG to anomalous galactic rotation 
velocities. CG has recently been fitted to rotation data for 138 
galaxies\cite{MAN97,MAO11,MAO12,OAM12,OAM15}. The CHM precludes 
existence of a massive Higgs particle, but conformal theory is found 
to be compatible with a compound gauge diboson $W_2$ of mass 125 
GeV\cite{NESM4}, consistent with the observed LHC 
resonance\cite{ATL12,CMS12}.
\revision{
Fits of CG and the CHM to observed galactic and cosmological data
do not require dark matter\cite{NESM5}.
}

\section{Dark matter}
\par When it became possible to measure
\revision{ 
orbital velocities
} 
in the outer 
reaches of galaxies, they were found systematically to exceed the
uniformly decreasing value implied by standard Einstein/Newton gravity.
The general functional form of $v(r)$ was observed to level off at
a characteristic radial acceleration $a_0\simeq 10^{-10}m/s^2$.
This led to the conjecture that standard gravity due to observed 
galactic mass was augmented by some additional gravitational source.
Since this source was not directly observed it was called dark matter.
\par Alternatively, general relativity might be modified to account
for this excess centripetal acceleration.  The most successful model 
assumes modified Newtonian dynamics (MOND)\cite{MIL83,FAM12}. 
The basic postulate for radial acceleration $a$, given Newtonian $a_N$, 
is that $a^2\to a_Na_0$ for $a_N \ll a_0$.
\par When conformal gravity (CG) is applied to a Schwarzschild model
(a central gravitational source with spherical symmetry)
it has an exact solution in the form of Schwarzschild radial potential
$B(r)$\cite{MAK89,MAN91}, for constants related by 
$\alpha^2=1-6\beta\gamma$\cite{MAN91}.  
Outside a source of finite radius\cite{MAK89},
\begin{eqnarray} \label{Brfn}
B(r)=-2\beta/r+\alpha+\gamma r-\kappa r^2.
\end{eqnarray}
$B(r)$ determines circular geodesics with orbital velocity $v$ such that
$v^2/c^2=ra/c^2=\half rB^\prime(r)=\beta/r+\half\gamma r-\kappa r^2$.
The Kepler formula is $ra_N/c^2=\beta/r$, from a 2nd order equation.
The 4th order conformal equation introduces two additional constants
of motion, radial acceleration $\gamma$ and cutoff parameter $\kappa$, 
which determines the radius of a galactic halo\cite{NESM3}. 
Classical gravitation is retained at subgalactic distances be setting
$\beta=Gm/c^2$ for a spherical source of mass $m$\cite{MAN06}.
\revision{
\par The depleted halo model\cite{NESM3}, described below, treats
all matter outside a defined galactic radius as uniform and isotropic.
Only spherical symmetry is considered. 
Following Mannheim and Kazanas\cite{MAK94}, galactic mass within this
radius is treated essentially by classical gravitation, describing
detailed nonspherical geometric structure.  Dark matter is replaced
by the anomalous acceleration parameter $\gamma$\cite{MAK89}.
}
\par An alternative to the multiplicative postulate of MOND is provided 
by additive acceleration parameter $\gamma$, which has the conceptual
advantage of arising from a well-defined variational field theory.
Fits to anomalous galactic rotational velocities by CG and MOND
are of comparable accuracy in the flat velocity range.  However,
the halo cutoff parameter $\kappa$, unique to CG, is found to be 
relevant at very large galactic radii\cite{MAO11,MAO12,NESM7}.

\section{Dark energy and Hubble expansion}
\par The Higgs scalar field\cite{HIG64,CAG98} is an essential element
of electroweak physics.  It has a spontaneously generated finite
amplitude, constant in spacetime, responsible for finite mass of
gauge bosons and fermions.   Retaining Higgs
$V(\Phi^\dag\Phi)=-(w^2-\lambda\Phi^\dag\Phi)\Phi^\dag\Phi$, which
depends on two assumed constants $w^2$ and $\lambda$\cite{HIG64,CAG98},  
the postulate of universal conformal symmetry\cite{NES13} requires
the CHM Higgs Lagrangian density to acquire a gravitational term, 
$-\sxth R\Phi^\dag\Phi$\cite{MAN06}, where
$R=g_{\mu\nu}R^{\mu\nu}$, trace of the Ricci tensor. 
The variation of Ricci  R on a cosmic time scale implies a very 
small but universal source density for the $Z_\mu$ neutral gauge field.
Dressing of the scalar field by $Z_\mu$ determines Higgs parameter 
$w^2$ and dressing by diboson $W_2$ determines $\lambda$\cite{NESM4}.
These two parameters and Ricci scalar $R$ imply finite $\Phi$ amplitude
and broken gauge and conformal symmetry.
\par In the uniform, isotropic cosmic geometry assumed for cosmology, 
the CHM implies a Friedmann cosmic evolution equation\cite{NES13,NESM2} 
with parameters determined by the scalar Higgs field.  This modified 
Friedmann equation contains an effective cosmological constant, 
defining dark energy density.  The integrated luminosity 
distance, computed as a function of redshift, fits observed data 
back to the CMB (cosmic microwave background)\cite{NESM1}.  Omitting 
cosmic mass and curvature, the fit to observed Hubble expansion data,
with centrifugal acceleration, is accurate back to redshift 
$z=1$(7.33Gyr)\cite{NESM2}.

\section{Depleted halo model}
\par CG and the CHM are consistent but interdependent\cite{NESM5} 
in the context of a depleted dark halo model \cite{NESM3} for an 
isolated galaxy.  A galaxy of mass $M$ is modeled by spherically
averaged mass density $\rho_G/c^2$ within an effective galactic radius
$r_G$, formed by condensation of primordial uniform, isotropic matter 
of uniform mass density $\rho_m/c^2$. 
\revision{
A model valid for nonclassical gravitation can take advantage of spherical 
symmetry at large galactic radii, assuming classical gravitation within 
$r_G$.  Nonspherical gravitation is neglected outside $r_G$.  
The dark halo inferred from gravitational lensing and centripetal 
acceleration is identified with the resulting depleted sphere 
of large radius $r_H$\cite{NESM3}. 
}
Given mean mass density
${\bar\rho}_G/c^2$ within $r_G$, this implies empty halo radius
$r_H=r_G({\bar\rho}_G/\rho_m)^\thrd$.
\par CG determines source-free Schwarzschild potential $B(r)$ as
Eq.(\ref{Brfn}) outside galactic radius $r_G$\cite{MAK89,MAN06}.
As shown in detail below,  
the physically relevant particular solution for $B(r)$\cite{NESM5}
incorporates nonclassical radial acceleration $\gamma$ as a free 
parameter.  Its value is determined by the halo model.  Gravitational 
lensing by a spherical halo is observed as centripetal deflection of 
a photon geodesic passing from the external intergalactic space with 
postulated universal isotropic mass-energy density $\rho_m$ into the 
empty halo sphere. The conformal Friedmann cosmic evolution equation 
implies dimensionless cosmic acceleration parameters $\Omega_q(\rho)$
\cite{NESM3} which are locally constant but differ across the halo 
boundary $r_H$.  Smooth evolution of the cosmos implies observable 
centripetal particle acceleration $\gamma$ within $r_H$ proportional 
to $\Omega_q(in)-\Omega_q(out)=\Omega_q(0)-\Omega_q(\rho_m)$.  Uniform 
cosmological $\rho_m$ implies constant $\gamma$ for $r\leq r_H$,   
independent of galactic mass \cite{NESM5}.  This surprising result is     
consistent with recent observations of galactic rotational velocities
for galaxies with directly measured mass\cite{MLS16,NESM7}, implying
that radial acceleration $a$ observed as orbital velocity is a function
of Newtonian $a_N$ independently of orbital radius and galactic mass.
\par In the CHM, observed nonclassical gravitational acceleration 
$\half\gamma c^2$ in the halo is proportional to\cite{NESM3} 
$\Delta\Omega_q=\Omega_q(0)-\Omega_q(\rho_m)=
\Omega_m(\rho_m)$, where, given $\rho_m$ and $H_0$, 
$\Omega_m(\rho_m)=
\frac{2}{3}\frac{{\bar\tau}c^2\rho_m}{H^2_0}$\cite{NESM1}, 
for Hubble constant $H_0$ and ${\bar\tau}<0$.
Thus the halo model determines constant $\gamma$ from
uniform universal cosmic baryonic mass density $\rho_m/c^2$,
which includes radiation energy density here.

\section{Consistency of CG and CHM}
\par CG and CHM must be consistent for an isolated galaxy and its
dark halo, observed by gravitational lensing.  CG is valid for anomalous
outer galactic rotation velocities in the static spherical Schwarzschild
metric, solving a differential equation for Schwarzschild gravitational
potential $B(r)$ \cite{MAK89,MAN06}.  The CHM is valid for cosmic
Hubble expansion in the uniform, isotropic FLRW metric, solving a
differential equation for Friedmann scale factor $a(t)$ \cite{NESM1}.
Concurrent validity is achieved by introducing a common hybrid metric

\begin{eqnarray} \label{xmet}
ds^2=g_{\mu\nu}dx^\mu dx^\nu=
-B(r)dt^2+a^2(t)(\frac{dr^2}{B(r)}+r^2d\omega^2). 
\end{eqnarray}
\par Metric tensor $g_{\mu\nu}$ is determined by conformal field 
equations derived from ${\cal L}_g+{\cal L}_\Phi$ \cite{NESM3},
driven by energy-momentum tensor $\Theta_m^{\mu\nu}$, where
subscript $m$ refers to conventional matter and radiation.
The gravitational field equation within halo radius $r_H$ is
\begin{eqnarray}
X_g^{\mu\nu}+X_\Phi^{\mu\nu}=\half\Theta_m^{\mu\nu},
\end{eqnarray}
where $X^{\mu\nu}$ is a metric functional derivative\cite{MAN06,NESM5}.
The gravitational equations are decoupled by
separating mass/energy source density $\rho$ into uniform isotropic
mean density ${\bar\rho}$ and residual ${\hat\rho}=\rho-{\bar\rho}$,
which extends only to galactic radius $r_G$ and integrates to zero
over the defining volume.
Defining mean density ${\bar\rho}_G$ and
residual density ${\hat\rho}_G=\rho_G-{\bar\rho}_G$, and assuming
$\Theta_m^{\mu\nu}(\rho)\simeq
 \Theta_m^{\mu\nu}({\bar\rho})+\Theta_m^{\mu\nu}({\hat\rho})$,
solutions for $r\leq r_G$ of the two equations
\begin{eqnarray} \label{Twoeqs}
X_g^{\mu\nu}=\half\Theta_m^{\mu\nu}({\hat\rho}_G),
X_\Phi^{\mu\nu}=\half\Theta_m^{\mu\nu}({\bar\rho}_G)
\end{eqnarray}
decouple, implying a solution of the full equation.  

\section{Computed parameters of Schwarzschild potential B(r)}
\par Given mass/energy source density $f(r)$ enclosed within $\R$,
the Schwarzschild field equation is \cite{MAK89,MAN91}
\begin{eqnarray} \label{Beq}
 \dar^4(rB(r))=rf(r), 
\end{eqnarray}
for $f(r)\sim(\Theta_0^0-\Theta_r^r)_m$ determined by source 
energy-momentum tensor $\Theta^{\mu\nu}_m$\cite{MAN06}.
\par Derivative functions 
$y_i(r) =\partial^i_r(rB(r))$ for $0\leq i\leq3$
satisfy differential equations\cite{MAN06,NESM5} 
\begin{eqnarray}
 \dar y_i=y_{i+1},0\leq i\leq2, \nonumber\\
 \dar y_3=rf(r) .
\end{eqnarray}
The general solution, for
independent constants $c_i=y_i(0)$, determines coefficients
$\beta,\alpha,\gamma,\kappa$ such that at endpoint $\R$
\begin{eqnarray}
 y_0(\R)=-2\beta+\alpha \R+\gamma \R^2-\kappa \R^3,\nonumber\\
 y_1(\R)=\alpha+2\gamma \R-3\kappa \R^2,\nonumber\\
 y_2(\R)=2\gamma-6\kappa \R,\nonumber\\
 y_3(\R)=-6\kappa .
\end{eqnarray}
Gravitational potential $B(r)$ is required to be
differentiable and free of singularities. 
$c_0=0$ prevents a singularity at the origin.  Specific values of 
$\gamma$ and $\kappa$, consistent with Hubble 
expansion and the observed galactic dark halo \cite{NESM1,NESM3}, 
can be fitted by adjusting $c_1,c_2,c_3$, subject to 
$c_0=0,\alpha^2=1-6\beta\gamma$\cite{MAN91}.
\par A particular solution for $B(r)$ \cite{MAK89,MAN91}, assumed by 
subsequent authors, derives an integral for $\gamma$ that vanishes for 
residual source density ${\hat\rho}$.  This is replaced here by an 
alternative solution for which $\gamma$ is a free parameter\cite{NESM5}.
Because the Weyl tensor vanishes identically in uniform geometry,
CG applies only to residual density ${\hat\rho}$.
\par The proposed particular solution, given $\gamma,\kappa$, is
\begin{eqnarray} \label{rBeq2} 
 rB(r)=y_0(r)=-\sxth\int_0^rq^4fdq+\alpha r-\half r\int_r^\R q^3fdq
\nonumber\\
 +\gamma r^2+\half r^2\int_r^\R q^2fdq
 -\kappa r^3-\sxth r^3\int_r^\R qfdq .
\end{eqnarray}
Integrated parameters $c_i=y_i(0)$ are
$c_1=\alpha-\half \int_0^\R q^3fdq$,
$c_2=2\gamma+\int_0^\R q^2fdq$,
$c_3=-6\kappa-\int_0^\R qfdq$, and at $r=\R$,
$2\beta=\sxth\int_0^\R q^4fdq$.
Term $\gamma r^2+\half r^2\int_r^\R q^2fdq$ in this solution
differs from prior reference\cite{MAK89}.  Here $\gamma$ is a free
parameter that determines generally nonzero $c_2$. 
\par For an isolated single spherical solar mass in a galactic halo,
mean internal mass density ${\bar\rho}_\odot$ within $r_\odot$
determines an exact solution of the conformal Higgs gravitational
equation, giving internal acceleration 
$\Omega_q({\bar\rho}_\odot)$.
Given $\gamma$ outside $r_\odot$,
continuous acceleration across boundary $r_\odot$,
\begin{eqnarray}  
\half\gamma_{\odot,in}c^2-cH_0\Omega_q({\bar\rho}_\odot)=
 \half\gamma c^2-cH_0\Omega_q(0),
\end{eqnarray}
determines constant $\gamma_{\odot,in}$ valid inside $r_\odot$.
$\gamma_{\odot,in}$ is determined by local mean source density
${\bar\rho}_\odot$. $\gamma$ in the halo is not changed.  Its value is 
a constant of integration that cannot vary in the source-free 
halo\cite{NESM3,NESM5}.  Hence there is no way to determine
a mass-dependent increment to $\gamma$.  This replaces the usually 
assumed $\gamma=\gamma_0+N^*\gamma^*$ by $\gamma=\gamma_H$, 
determined at halo boundary $r_H$.

\section{Implications for cosmology}
\par The common assumption for galactic growth is that a primordially
accumulated dark matter halo subsequently attracts baryonic matter
to form an observable galaxy.  
\revision{
Conformal theory as well as MOND reverse this sequence,
while eliminating the need for dark matter.
The nonclassical CG gravitational acceleration is a byproduct of
the gravitational accumulation of baryonic matter attracted to a 
growing galaxy. The CHM generates a uniform constant nonclassical 
centripetal acceleration within a halo of large expanding radius. 
The rate of galactic growth must depend on the net incoming flux of    
matter diffusing across the halo boundary, where the 
net gravitational radial acceleration vanishes. 
}
\par Galactic collision is initiated by halo contact.  Because halo 
volume is determined by galactic mass, it must remain constant, 
implying distortion of colliding halo boundaries analogous to 
collision of two spherical balloons.  A dynamical model must
consider diffusion of matter across a changing halo boundary. 
A new observable phenomenon affects neighboring galaxies in a
galactic cluster.  Once halos are in contact, the accessible source
of primordial matter is restricted, hence reducing the rate of growth
of both colliding galaxies.  This should be observed as cessation   
of growth from the primordial background in the extreme case of a 
galaxy completely surrounded by a cluster of contiguous halos. 
\revision{
\par The empirical correlation relation of McGaugh et al\cite{MLS16}
establishes total radial acceleration as a function of its baryonic
Newtonian value.  This implies CG $\gamma$ independent of galactic 
mass\cite{NESM7}, which places a strong constraint on galactic rotation 
curves. The selected particular solution, Eq.(\ref{rBeq2}) for $B(R)$, 
which differs from\cite{MAK89}, depends on independently determined 
$\gamma$.  Resulting nonzero $c_2$ implies singular Ricci scalar at the
galactic center\cite{NESM5}, relevant to formation of a supermassive
black hole. 
}
\par Because the coefficient of the source term in the conformal
Friedman equation is negative, primordial energy density must cause 
centrifugal acceleration.  This may create a dynamical Big Bang 
in the CHM without requiring a separate field.  The relevancy of CHM 
should be explored.
\revision{
Weak time-dependence of scalar field $\Phi$ is implied\cite{NESM1,KH16}.
For large redshifts, the Friedmann equation for scale factor $a(t)$
and the CHM equation for $\Phi(t)$ must be integrated together. 
A changed Higgs amplitude $\phi_0$ affects initial atomic abundances.
} 
\vfill\eject

\vfill\eject
\end{document}